# Code Obfuscation for the C/C++ Language

A dissertation submitted in partial fulfilment of

the requirements for the degree of

BACHELOR OF SCIENCE in Computer Science

in

The Queen's University of Belfast

by

Dominik Picheta

Wednesday, 02 May 2018

**SCHOOL OF ELECTRONICS, ELECTRICAL ENGINEERING and COMPUTER SCIENCE**

**CSC3002 - COMPUTER SCIENCE PROJECT**

**Dissertation Cover Sheet**

A signed and completed cover sheet must accompany the submission of the Software Engineering dissertation submitted for assessment.

Work submitted without a cover sheet will **NOT** be marked.

Student Name: Dominik Picheta                    Student Number: 40122251

Project Title: Code Obfuscation for the C/C++ Language

Supervisor: Dr. Stuart Ferguson

# Contents

# 1    Acknowledgments

Thanks to Dr. Stuart Ferguson for supervising this fascinating project and for offering guidance and help when required.

# 2    Abstract

Obfuscation is the action of making something unintelligible. In software development, this action can be applied to source code or binary applications. The aim of this dissertation was to implement a tool for the obfuscation of C and C++ source code. The motivation was to allow proprietary code to be distributed to third-parties without risking a recreation of the intellectual property within it. While many obfuscators exist, they seldom focus on software that is distributed in source code form. This dissertation presents the challenges and successes that arose during the development of a C and C++ source code obfuscator using the Nim programming language [1].

# 3    Introduction and Problem Specification

There are three major types of attacks that software protection mechanisms aim to defend against, these attacks can be classified as software piracy, reverse engineering, and tampering. Collberg [2] defines these attack types and suggests tools to defend against each of them. For the purposes of this dissertation, the focus will be on reverse engineering protection.

The main defense against reverse engineering is obfuscation [2]. Obfuscation can be said to protect the intellectual property (IP) of software from reverse-engineering attacks. It is important to protect the IP as it can include sensitive data, algorithms, or the design of the software which the developer may not wish to be copied [3, Sec. 2]. Obfuscation is defined as the transformation of code into something unintelligible which preserves the semantics of the original code.

Software can be distributed in two forms, as source code or compiled machine code. Machine code instructions can be executed directly by a computer's central processing unit (CPU). For a programmer, reading these instructions is not an easy task and writing them is even more difficult. Because of this machine code is usually generated from source code by a special program called a compiler. An executable or binary file stores machine code instructions in a format that is specific to a particular operating system.

Special applications called decompilers can be used to reverse the process of compilation. That is, they take an executable file as input and output high-level source code which matches the functionality of the executable. A disassembler is a special kind of decompiler which translates machine language into assembly language. Decompilers targeting languages such as C++ will often use a disassembler as the first stage of the decompilation process.

Decompilers are an important and often used tool for the reverse engineering of machine code. This is why applications designed for obfuscating executables focus on the obstruction of disassembly and decompilation [4, Sec. 1]. The aptly named `obfuscator` application developed in this dissertation does not put any effort into obstructing decompilation. Instead it focuses on making the source code as difficult to reverse engineer as possible, by decreasing its overall comprehensibility.

Reverse engineering source code is all about understanding it, figuring out the control flow and how it interacts with its input data. With enough time, every piece of code can be reverse engineered, so it is impossible to guarantee complete safety [5, Sec. 1]. But making the reverse engineering economically impractical is often enough.

A tool that obfuscates source code is incredibly valuable for preventing these attacks. It can be a cheap way to protect IP from third parties who may have disassembled an executable, or those who need to have access to the source code in order to build it on a niche platform [6].

## 3.1  Challenges

The actual act of obfuscating source code is no easy task. There are many distinct challenges that need to be explored.

### 3.1.1  Parsing

Parsing is a necessary pre-requisite for obfuscation, since an obfuscator needs to understand the code it is obfuscating. Programming languages are often difficult to parse correctly, this is especially the case for C++ which is often considered to be the most difficult language to parse. The reason is that C++ has a grammar which is both huge and ambiguous, with many new revisions such as C++20 still being designed and implemented.

```cpp
void f(double adouble) {
  int i(int(adouble));
}
```

Listing 1: Most vexing parse: an example of ambiguity in C++.

Because of the complexity involved in parsing C and C++ code, developing a full-featured parser would take a lot of time. It is possible to take some shortcuts, perhaps a minimal parser that picks out desired syntactic elements of a source code file could be written quickly, but it would always be missing vital features for some users. In such a parser the foundations would always be rough and without taking into account the full grammar of the C++ language from the start it would be doomed to a dead end at some point.

Thankfully the most popular C and C++ compilers, `gcc` and `clang`, are open source. Even though their codebase is large, the code responsible for parsing is logically separated from any other functionality, making it reusable from other applications. The compiler's parsers are incredibly robust and support every current C/C++ feature.

Reusing a parser comes with its own challenges, but solving them is much easier than writing a custom parser from scratch. Section 6 discusses the implementation challenges in detail.

### 3.1.2 Transformations

Once the code can be parsed, it needs to be represented in the program's memory. The representation needs to be flexible enough to be mutable, in order to facilitate various code transformations. The transformations are necessary to obfuscate the code.

The objective is to apply transformations that deliberately obfuscate the source code of a program, so that its purpose or logic is concealed without any alterations being made to its functionality.

Transformations can be separated into three main classes [7, Sec. 2]:

1. *Data transformation*: The data inside a program is usually presented in a way that makes it easily readable to programmers. At the most fundamental level this applies to literals such as integers and strings. An integer can be represented using multiple different radices including hexadecimal, octal or decimal. An integer can also be transformed into a constant expression (a form of *data encoding* obfuscation [8, Sec. 3.3.2]), and other static data may be converted to procedural data as well [7, Sec. 2]. In most languages strings can be escaped using various radices as well, including hexadecimal escape sequences [6]. More advanced forms of data transformations include variable splitting and changing the scope of variables [6].

2. *Lexical transformation*[1] [2]: The code's lexical information refers to the names of identifiers, the comments, and the whitespace used to indent the code to make it more readable. When performing this transformation, comments and whitespace are usually completely removed [8, Sec. 3.3.1], but bogus whitespace or comments designed to obfuscate the code can be introduced instead. A more complicated transformation involves changing the identifiers in the code, including the names of variables, classes, and functions. Identifiers can be renamed in multiple ways as long as the names are unique, they can be changed randomly, based on an algorithm such as a `hash` function, or by look up in a look-up table. This transformation does not modify any of the semantics in the code.

3. *Control transformation*: The semantics of the program are modified through this transformation. As a result, it may affect the runtime of the program in a negative way [7]. Some techniques that can be used for this transformation include addition of dead code or the modification of control structure predicates

---

[1]Otherwise known as *layout transformation* [7], [9]

such as if statements or for loops to make them more difficult to understand [5, Sec. 1].

Researchers [5], [7], [10], [11] are always investigating new and more complex ways to obfuscate code by coming up with novel transformation techniques. Implementing all such techniques is outside the scope of this dissertation, but the `obfuscator` project does offer a great modern test bed for them.

### 3.1.3 Rendering

The data structure that contains the obfuscated code doesn't reflect the code itself, so it cannot be easily written to a file. This data structure needs to be converted into a valid C or C++ source code representation.

The generated C or C++ source code needs to be free of errors. It cannot omit any syntactical constructs as that would prevent the code from being compiled for testing.

### 3.1.4 Correctness

An obfuscator is said to be correct if the obfuscated code it outputs is *always* functionally equivalent to the original code. Unfortunately an obfuscator has a very real chance of changing the semantics of code in a way that breaks the resulting application. This should never happen, but as with all software, bugs are possible and will need to be fixed.

The correctness of an obfuscator is important, but proving this correctness isn't easy. Changes in semantics may be subtle, they may only show up on certain niche platforms or under very specific runtime circumstances. For a language like C or C++ verifying the correctness thoroughly is practically impossible, but a good testing methodology can find many issues.

C and C++ are both compiled languages, they use a compiler that performs static analysis of the code to verify certain aspects of its correctness. This is the first correctness test, if the obfuscator generates invalid code the compilation will often fail with an error. It's important to verify that the obfuscated code compiles. Sometimes the compiler may give useful warnings as well, it is also a good idea to check that no new warnings have been introduced in the obfuscated code.

The second correctness test is a runtime test. The resulting executable of the obfuscated program should be tested. This is challenging as testing certain pieces of software thoroughly may not be possible. Even if rigorous testing cannot be done, a best effort should be made to ensure the software works as intended after its source code has been obfuscated. Test suites provided by the software should be used whenever possible to make the testing easier and more reliable.

### 3.1.5 Quality of obscurity

Collberg [2] explains that an obfuscator should maximize the *obscurity* of the obfuscated code. The obscurity of code refers to how time-consuming understanding and reverse engineering it is.

Unfortunately measuring obscurity empirically is difficult, it would require a controlled experiment involving professional developers and a measurement of the time it takes them to understand an obfuscated vs. an original piece of code. As an example Regano et al. [12] has performed such an experiment and found that their `VarMerge` obfuscator "reduces by six times the number of successful attacks per unit of time."

Performing a similar experiment would require significant amount of resources and time which are not available for this dissertation. As an alternative, there are ways to measure obscurity indirectly.

The alternative way to measure obscurity is by calculating the complexity of code. There are multiple metrics which calculate this:

- *McCabe's cyclomatic complexity* which measures the number of independent control paths in a program.
- *Halstead complexity measures* which combine a number of properties including the number of distinct operators and operands.

Naeem et. al. [11] investigates these metrics in the context of decompilers and obfuscators. Some alternatives to McCabe's and Halstead's metrics are offered, including a measure of the program size, the conditional complexity and the identifier complexity.

The advantages of McCabe and Halstead metrics are that they are widely supported and tools exist for measuring them. László et. al. [5] uses McCabe's complexity together with program size for evaluating the obscurity of their obfuscator. Unfortunately these metrics are not a good way to evaluate the obscurity of the obfuscator developed in this dissertation. They measure a very specific complexity feature which is not affected by the `obfuscator` application.

A related concept is that of *resilience*, i.e. an obfuscator is resilient if it confuses an automatic de-obfuscator. A developer wishing to reverse engineer obfuscated source code will often turn to a tool that "prettifies" or "auto-formats" the code. These tools might be able to make reverse engineering significantly easier.

For this dissertation, a proposed approach to testing the quality of the obfuscator is to feed its output to a "prettifier" tool such as `clang-format`[2]. A `diff` tool can then be used to measure the difference between the obfuscated code after it's been prettified and the original code.

This approach is novel and untested but has a good chance of evaluating both the resilience and the obscurity in the context of the `obfuscator` application and how it compares to other source-to-source obfuscators.

## 3.2 Related work

Obfuscating C and C++ code is a popular task that started out as a hobby in the 80's, thanks to the International Obfuscated C Code Contest (IOCCC)[3] [7, Sec. 2.C]. In the case of the IOCCC, obfuscation is a creative exercise performed mainly by human programmers for the purposes of entertainment. In later years, software was developed to systematically obfuscate code as a defense against reverse engineering.

Today many commercial and open source C/C++ obfuscators exist, including Stunnix[4], COBF[5], CShroud[6] and Tigress[7].

At the time of writing a Stunnix single developer license is priced at \$449, the feature set of the software goes far beyond obfuscation so the price tag isn't without merit. As far as transformations go, the Stunnix obfuscator performs standard layout and data obfuscations.

CShroud is an open source program licensed under the GPL, it performs layout transformations including removal of comments and indentation, in addition it converts control structures such as `for`, `while`, `do/while`, `if/else` and `switch` into `if/goto` structures [8, Sec. 1.4]. Unfortunately CShroud doesn't appear to be maintained anymore.

---

[2] https://clang.llvm.org/docs/ClangFormat.html
[3] http://ioccc.org/
[4] http://stunnix.com/prod/cxxo/
[5] https://www.plexaure.de/cms/index.php?id=cobf
[6] https://sourceforge.net/projects/cshroud/
[7] http://tigress.cs.arizona.edu/

Tigress is another obfuscator, unlike CShroud its latest release is recent and appears to be maintained at the time of writing. The Tigress website describes it as "a diversifying virtualizer/obfuscator for the C language that supports many novel defenses against both static and dynamic reverse engineering and de-virtualization attacks."

Obfuscators that work on binary files also exist, one example of such an application is called `obfuscator-llvm`[8] which can output an obfuscated binary code file [9, Sec. 2.1.2]. This is different to the obfuscators mentioned above which will output an obfuscated source code file.

Obfuscation is also a healthy subject of study. Research papers often investigate the ideal transformations that can be applied to achieve the best defense against reverse engineering.

# 4   System Requirements Specification

This section is going to specify the requirements, assumptions and constraints of the `obfuscator` application developed as part of this dissertation.

## 4.1   User interaction

The end user will interact with the `obfuscator` application using a command-line interface. This is in line with how most software developers interact with C and C++ compilers.

The application will support a number of flags to modify its behaviour. The usage will be as follows:

- `./obfuscator <filename>.c` or `./obfuscator <filename>.cpp`

  - Parses the source code located at the specified filename, obfuscates it and saves the obfuscated source code in `<filename>.obf.c` or `<filename>.obf.cpp`.

- Command-line flags:

  - `-o:<outputFile>`

---

[8]https://github.com/obfuscator-llvm/obfuscator/wiki

          ∗ Save the obfuscated source file in the file path specified.
- `--rename`
  - ∗ Rename identifiers in the source code (disabled by default).
- `-I:<includePath>`
  - ∗ Specify an include header path for the C/C++ source code parser.

## 4.2   Assumptions

The `obfuscator` application will be given syntactically and semantically valid C or C++ code. Validity will be defined in terms of the standards supported, which will include C99 [13] and C++98 [14]. Some features of the C11 and C++11 standards may also be supported, but code containing those features may be viewed as invalid by the `obfuscator` application.

A relatively modern computer system with an access to a terminal will be required to run the application. Due to the CLI nature of the application the end user will need to be comfortable with a terminal to use the obfuscator.

## 4.3   Constraints

The initial version of the `obfuscator` application will require macOS to run. This is mainly a constraint due to the operating system that development took place on, there is no reason the application cannot be compiled on other operating systems and platforms with little to no changes to the code.

## 4.4   Requirements

At a high level, the `obfuscator` application is expected to parse a single C or C++ source code file, obfuscate it and save the obfuscated code into a new file. The requirements including the specific obfuscation transformations that should be applied to the code are outlined below.

### 4.4.1   Formatting and comments

Most developers take great care to format their code in a logical and readable manner. They also add useful comments to code which describes its semantics. Removing formatting and comments is a great way to obfuscate code.

The obfuscator should perform a *lexical transformation* to remove all comments and formatting including whitespace and newline characters from the code. Listing 2 and Listing 3 shows the transformation in action.

```c
int main () {
  helloWorld(42); // Say hello.
}
```

```c
int main(){helloWorld(42);}
```

Listing 2: Original code                        Listing 3: Transformed code

The transformation needs to be aware of C syntax. Removing all whitespace indiscriminately from C or C++ source code would lead to errors. This is evident in Listing 3 where whitespace is required to separate the function's return type from the `main` identifier.

It is also important to place semi-colons in the correct positions. The end of each statement needs to be delimited by a semi-colon, other syntactic elements have special rules for the placement of this delimiter, including predicates in `for` loops. These need to be handled appropriately.

### 4.4.2   Integer and string literals

Literals are present in almost all source code. Integer and string literals can be obfuscated in a relatively simple manner through *data transformation*.

Understanding integer literals is easiest when they are represented in the commonly used decimal numeral system. In C and C++ an integer literal can also be represented in hexadecimal or octal. For the purposes of obfuscation all integer literals should be represented using hexadecimal. This transformation is shown in Listing 4 and Listing 5.

```c
int main () {
  int age = 23;
}
```

```c
int main () {
  int age = 0x17;
}
```

Listing 4: Original code                        Listing 5: Transformed code

The obfuscation of integer literals should go a step further. Each integer literal should be transformed into a constant expression that reproduces the original literal. A single literal should be separated into at least 4 different random integers, these should be added and subtracted together to give the original literal. This transformation is shown in Listing 6 and Listing 7.

```c
int main () {
  int age = 23;
}
```

Listing 6: Original code

```c
int main () {
  int age = 759 + 78 - 826 + 12;
}
```

Listing 7: Transformed code

String literals in source code are usually displayed in a human readable format. Every string should be obfuscated by randomizing its representation, each character should be represented either using a hexadecimal escape sequence, a decimal escape sequence or as-is. The mixing of different escape sequence formats should confuse the reader and ensure that a simple tool cannot be written to deobfuscate it. This transformation is shown in Listing 8 and Listing 9.

```c
"Hello"
```

Listing 8: Original code

```c
"\x48""e\154l\x6F"
```

Listing 9: Transformed code

To ensure syntax compatibility with C and C++, the transformation must start a new string literal after each hexadecimal escape sequence. Otherwise code like `"\x48e"` would be incorrectly interpreted as a single `48E` hexadecimal value instead of two `\x48` and `e` characters.

### 4.4.3   Boolean expressions

The control paths in source code are essential to its execution, as a result they are an important way to learn about the code. Many constructs used to control the flow of execution contain boolean expressions in their predicates, including `if`, `for` and `while` statements. As a result transforming boolean expressions is an important method of obfuscating code.

The boolean expressions should be transformed in such a way as to yield the same results. Doing so is a form of *control transformation*. A simple transformation that should be applied is creating a copy of the original boolean expression, negating it and placing it inside a newly created `&&` (boolean `AND`) expression.

```
int x = 42;
if (x > 12) {}
```

Listing 10: Original code

```
int x = 42;
if ((x > 12) && !!(x > 12)) {}
```

Listing 11: Transformed code

This adds extra noise which, when applied to every boolean expression, significantly maximizes the obscurity of the obfuscated code.

### 4.4.4 Identifiers

It is often said in jest that there are only two hard things in Computer Science: cache invalidation and naming things [15]. This has a ring of truth to it. Software developers often take great care to name variables, functions and other identifiers in a way that makes the code easy to understand. Obfuscating these names will have a great effect on the comprehensibility of the source code.

Ideally all identifier types should be obfuscated, but variable and function names are a good start. The relevant transformation should be aware of the visibility of the identifier. If an identifier is exported via a header file, it should not be transformed as doing so may lead to errors due to cross-source dependencies. Identifiers that are not exported must be transformed. The transformation is described in Listing 12 and Listing 13.

```
int foo(int param) {
  int x = 42 + param;
  return x;
}

int main() {
  int y = 2;
  return foo(y);
}
```

Listing 12: Original code

```
int i_acbd(int i_eca0) {
  int i_9dd4 = 42 + i_eca0;
  return i_9dd4;
}

int main() { ❶
  int i_4152 = 2;
  return i_acbd(i_4152);
}
```

Listing 13: Transformed code

Note how in Listing 13 the main function defined at ❶ isn't obfuscated. This is because the C linker will look for this function name during compilation to use as the entry point of the program. Renaming it would cause an error.

The rest of the identifiers have all been obfuscated. How these identifiers are obfuscated is up to the `obfuscator` implementation. In the case of Listing 13 the first 4 characters of an MD5 hash function's output have been used.

### 4.4.5 Testing

An obfuscator lends itself well to testing, because while it's a complicated piece of software it can be easily executed using simple scripts. A `tester` program or script should be written to execute the `obfuscator` on sample source code and verify that the original and obfuscated source code compiles successfully.

The first testing category will be unit tests where sample source code containing a single category of specific C and C++ language features will be tested. For example, two unit tests can be created for array declaration syntax and if statements. A good range of unit tests should be created to ensure the obfuscator works as expected, and to make it easier to find issues during development.

The `tester` should also perform integration tests. The difference between these and unit tests is that they will test the ability to obfuscate a full project. For every project, every source code file should be obfuscated, compiled and tested. Real open source projects should be used for this purpose to ensure the `obfuscator` works effectively.

## 5   Design

The architecture of the `obfuscator` application will be modelled after a typical compiler. A compiler begins its work in the same manner as the obfuscator: by parsing source code. The output of these tools is the differentiator, an obfuscator will output obfuscated source code while a compiler will output machine code or an executable instead.

A high-level view of `obfuscator`'s architecture is shown in Figure 1.

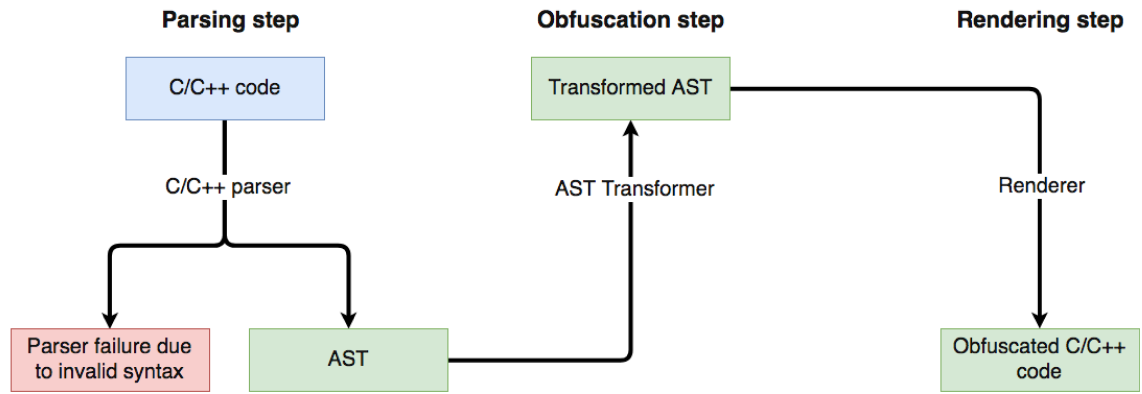

Figure 1: The `obfuscator` architecture

There are three distinct components in this architecture: the parser, AST transformer and renderer. These components are described in the following sub-sections.

## 5.1   Parser

The parser component is responsible for parsing valid C and C++ source code. The output of the parser should be a mutable Abstract Syntax Tree (AST) data structure.

An AST is a tree representation of the parsed source code. The tree itself has a structure that describes the source code and each node in the tree can store additional data about the syntactic construct it represents. For example, an integer literal node may contain the integer value as part of its data. Each node can also store 0 or more children. Figure 2 shows what an AST might look like.

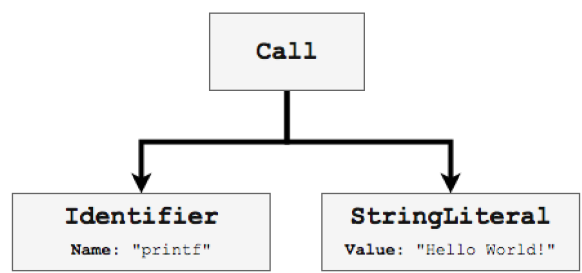

Figure 2: A simple AST for `printf("Hello World!");`

The fundamental piece of information about an AST node is its "kind". This information determines the type of syntax element the node represents, as an example

`IntLiteral` represents an integer literal and `IfStmt` represents an `if` statement. Each node kind contains different data and so needs to be handled differently. Figure 3 shows an annotated version of the AST shown in Figure 2.

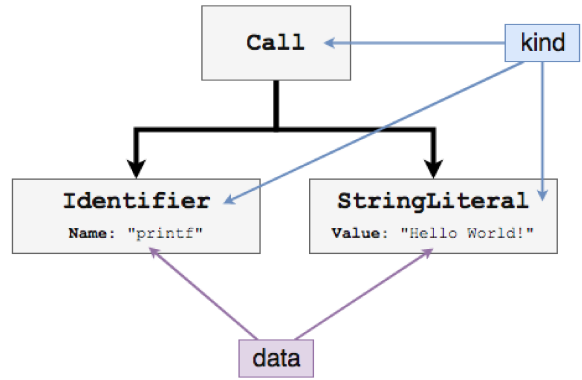

Figure 3: An annotated AST for `printf("Hello World!");`

The parser is expected to produce an AST for all possible valid source code inputs. For cases where the source code is incorrect, the parser should produce a good error message describing the problem with a line number pointing to its location.

As mentioned previously, a custom parser will not be written. Section 6 has more details.

## 5.2  Transformer

The transformer component is responsible for performing all the various transformations specified in section 4.

After the parser performs its work and creates an AST, the transformer steps in. The transformer searches for patterns in the AST recursively, once a pattern is found the AST node is modified in a manner that is consistent with the specific transformation being applied.

As an example, consider the identifier transformation described in subsubsection 4.4.4. Among many others the transformer will look for AST nodes representing variable definitions, when found it will determine their visibility. Only if the variable is not exported will the transformer obfuscate it. The obfuscation will be performed by modifying the AST node's data, typically a `name` field which contains the name of the defined variable.

## 5.3   Renderer

The renderer component is the final component in the `obfuscator` architecture. It is responsible for converting an AST into its C or C++ source code representation.

The sheer number of different kinds of AST nodes used to represent C and C++ source code means that rendering them all is a difficult process. Having to implement a fully featured renderer would mean that the obfuscator could not be tested quickly.

A good way to work around this issue is to store the original code in each AST node. That way, if the renderer cannot handle a specific AST node kind, the original code can be used instead. This has the advantage that the renderer can be developed incrementally, but means that while the renderer is unfinished certain code will remain unobfuscated.

To demonstrate what this means, let's look at a larger AST example:

```
Call (origCode: "printf(\"Hello World!\");")
  Identifier (origCode: "printf", name: "printf")
  StringLiteral (origCode: "\"Hello World!\"", value: "Hello World!")
IfStmt (origCode: "if (true) {\n printf(\"true\"); \n}")
  Branch
    Identifier (origCode: "true", name: "true")
    StmtList (origCode: "printf(\"true\");")
      Call (origCode: "printf(\"true\");")
        Identifier (origCode: "printf", name: "printf")
        StringLiteral (origCode: "\"true\"", value: "true")
```

Listing 14: A textual representation of a larger AST

Listing 14 shows the AST tree of Listing 15. The children are indented to show the parent node they belong to. Note the data fields contained in each node.

Assuming that a renderer is used which does not support if statements, Listing 16 shows what the obfuscated code will look like.

```
printf(100);

if (true) {
  printf(200);
}
```

```
printf(0x64);if (true) {
  printf(200); ❷
}
```

Listing 15: Source code for the AST in Listing 14

Listing 16: Code rendered without if statement support

Because the renderer does not know how to handle `if` statements, it simply reuses the `origCode` of the AST node. This leaves the integer literal in ❷ unobfuscated. Of course, the renderer does not add any newline characters to the output, so the original code is simply appended after the first call to `printf`.

It's also important to note that the renderer is responsible for rendering literals. The transformer does not change the representation of an integer literal in any way, it's up to the renderer to decide how to render it.

# 6 Implementation and Testing

A high-level diagram showing the implementation of the `obfuscator` architecture is shown in Figure 4.

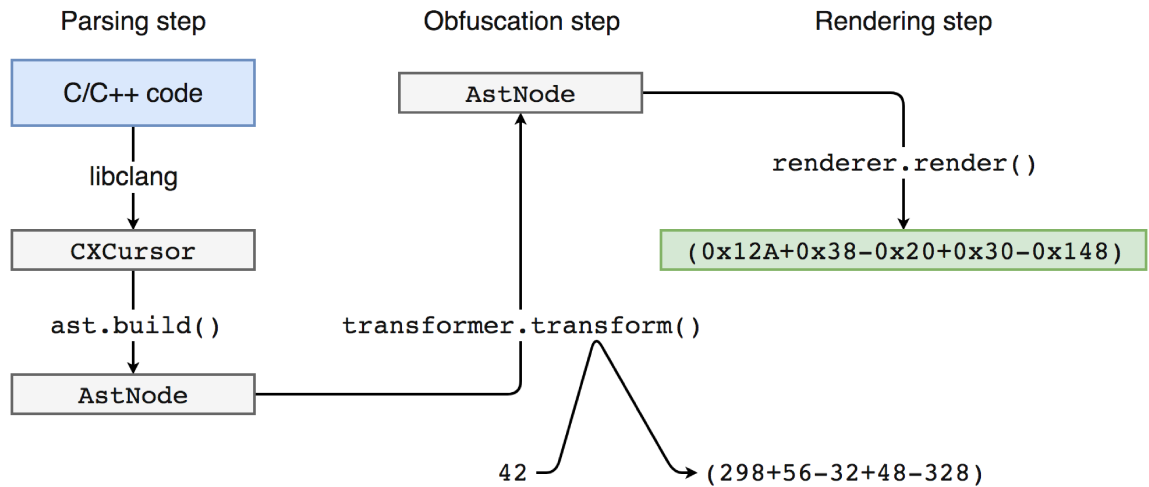

Figure 4: The `obfuscator` implementation

The following subsections describe the implementation in more detail. The source code of the implementation is available on the EEECS GitLab instance [1].

## 6.1   Programming language

The `obfuscator` application is implemented primarily in Nim[9]. Nim is a compiled systems programming language that compiles to C, C++ or Objective C. Nim produces executables that are dependency free, efficient and cross-platform, the language has some of the best metaprogramming features out there with support for things like procedural AST macros.

The choice of implementation language was made from the get go, the main reason for this choice was the author's familiarity with the language. As it turned out, the dependence on a C++ library meant that the only real choices were either C++ or Nim. Thanks to Nim's C++ compilation, interfacing with a C++ library is trivial, a feature that makes Nim just about the only practical alternative to C++.

Writing software in Nim is straightforward, C++ on the other hand is a colossal language with multiple edge cases and a memory model that makes writing software safely challenging. C++ also misses some modern language features, for example modules which are supported by Nim.

## 6.2   Parser

The `libclang` library is used to parse C and C++ source code. In order to make use of it, a thin Nim wrapper had to be written, this was made trivial thanks to the `c2nim` tool which generated most of it automatically.

The way in which the `libclang` library exposes information about the parsed source code is a bit unusual. It seems to have been designed for the purposes of Integrated Development Environment (IDE) introspection tools, which only need to understand a small subset of the source code and not modify it. The information is exposed via a `CXCursor` object which represents a single position in the source code, this object can be queried for information about the syntactical construct that is at the underlying cursor position.

This is a problematic API for two reasons:

- It's not immediately obvious what information about a particular `CXCursor` is available. The cursor's kind is exposed, but there is no exhaustive listing of queries that can be performed against each cursor kind. For example, there is no easy way to find out how to determine if a function call is an operator or not.

---

[9]https://nim-lang.org

- The `CXCursor` object cannot be modified. Even if it could be, the data about the syntactical construct is not exposed directly.

Research into these issues revealed more specific limitations of `libclang`, for example the inability to retrieve the value of an integer literal [16]. This was a showstopper for a while and it seemed like `libclang` would have to be abandoned in favour of something else.

Further investigation revealed that `libclang` is a C wrapper on top of the original clang parser written in C++. Using the clang parser directly, although much more difficult, was always a possibility. But looking at the libclang source code closely revealed certain abstract pointers to data exposed through the `CXCursor` object. A further look at how the different `libclang` query functions work revealed that these pointers are actually pointing to the original clang parser objects. By wrapping the underlying C++ classes in Nim, it is possible to access these objects and the information they store [17]. This approach allows the continued use of `libclang` and access to all the necessary information by falling back to C++ when necessary.

### 6.2.1 Mutability

As discussed in subsection 5.1, the parser component of the architecture needs to produce a mutable AST. The `CXCursor` object exposed by `libclang` doesn't support modifications, only queries. This is also true of the underlying clang parser objects.

A decision was made to define a custom AST object which would be built by iterating through all the `CXCursor` objects. The `build` function responsible for this is defined in the `ast` module and is used during the parsing step as shown in Figure 4.

The resulting `AstNode` object is a tree containing a representation of the parsed source code. Each `AstNode` object contains data specific to each AST kind. This data is stored inside the object's attributes so it's easy to know what information is stored about a particular AST kind.

The API exposed by the `ast` module is much more straightforward than that of `libclang` or even the clang parser itself. Because of that it might be worth exposing it as a library for other applications to take advantage of. Of course there is still plenty of room for improvement as it doesn't expose all of the information that the clang parser exposes.

### 6.2.2   Preprocessor

The C/C++ preprocessor is a completely separate tool to the parser. It is responsible for some important tasks, such as the inclusion of header files, macro expansions, and conditional compilation. Because this preprocessor is executed first, the parser exposes very little information about what the code looked like before preprocessing took place.

This creates multiple problems for the obfuscator, in general these problems mean that the obfuscator cannot restore some preprocessor directives. Some of these problems have been solved, but others persist. Table 1 shows a summary of preprocessor constructs and whether they are handled correctly by the `obfuscator` application.

Table 1: Summary of the preprocessor features `obfuscator` supports.

| Construct | Handled correctly | Example |
|---|---|---|
| Header inclusion | Yes | `#include <stdio.h>` |
| Macro expansion | Yes | Listing 17 |
| Conditional compilation | No | Listing 18 |

```
#define PI 3.14159

printf("%f", PI);
```

Listing 17: Macro expansion

```
#ifdef __unix__
# include <unistd.h>
#elif defined _WIN32
# include <windows.h>
#endif
```

Listing 18: Conditional compilation

The `libclang` parser provides information about inclusion directives and macro expansions. This allows the `obfuscator` to render these constructs in the obfuscated code.

Unfortunately no information is provided about conditional compilation directives. When parsed, the AST of Listing 18 contains only a single inclusion directive, with the included file depending on the operating system that the parser is executed on.

This is a limitation which causes the produced AST to always be dependent on the platform that it is produced on. For some use cases it is a serious limitation, but for others it may in fact be a feature. Solving this limitation is beyond the scope of this dissertation, but there are multiple approaches that can be considered in the future for solving it:

- Write a custom pre-processor which adds conditional compilation directives to the AST produced by `libclang` and the `ast` module.
- Investigate the `clang` source code to see if there is any possibility to enable some sort of parsing mode which will add the conditional compilation directives to the AST.

## 6.3  Transformer

The transformer component is implemented fully in Nim in just over 100 lines of code. This makes it the simplest component in the obfuscator.

This component's job is to transform the AST in such a way as to obfuscate it. The transformations performed include:

- Replacing integers literals with simple mathematical expressions. This is the transformation shown in Figure 4.
- Giving anonymous structs obfuscated names. This is used as an aid during rendering.
- Replacing the `==` and `!=` equality operators by `^` (XOR).
- Adding noise to boolean expressions, by adding a copy of the same expression with two boolean NOT prefixes, as described in subsubsection 4.4.3.
- Renaming function, variable, and function argument identifiers.

Most of these transformations are relatively trivial. The identifier renaming, which is most complicated, is described in more detail in the following section.

### 6.3.1  Renaming identifiers

The transformation itself is very simple, but deciding whether it should be performed for a specific identifier is non-trivial.

There are 3 different pieces of information collected by the `ast` module for the purposes of this transformation:

- *Unified Symbol Resolution (USR)*: this is a string that uniquely identifies functions, classes, variables, etc. across different files. It allows the obfuscator to change the names of variables consistently across different C/C++ source code files.

- *isGlobal flag*: this determines whether an identifier is defined inside any of the included header files.
- *referencedLocally flag*: this determines whether a referenced identifier is defined inside the file being obfuscated.

Global identifiers, as determined by the `isGlobal` flag, are not renamed because they may be used by external software. Similarly, only identifiers referenced locally are renamed.

The new name for identifiers is generated using the USR string. This string is hashed using MD5 and used as the new name. This produces consistent and reliable identifier obfuscations.

## 6.4   Renderer

Sometimes parser libraries implement their own rendering functionality. This is the case with `clang`, but this functionality cannot be reused because the obfuscator implements a custom AST.

A custom renderer is difficult to write, but it does give far more control about the code that is created. For an obfuscator this is really important.

The `obfuscator` application's renderer is written completely in Nim. As of writing, it supports a vast majority of syntactical constructs, with unsupported constructs being rendered using the original code as described in subsection 5.3.

The renderer intentionally omits whitespace as much as possible, this gets rid of all formatting and has the effect that there is no newline characters in the obfuscated code. The renderer is also responsible for rendering literals, it does so consistent with the system specification described in section 4.

## 6.5   Testing

The test suite is implemented in the `tester.nim` file. Running `nimble test` in the obfuscator's directory will compile the `tester` module and then execute it. Upon execution `tester` runs a suite of unit and integration tests. Figure 5 shows the successful execution of `nimble test`.

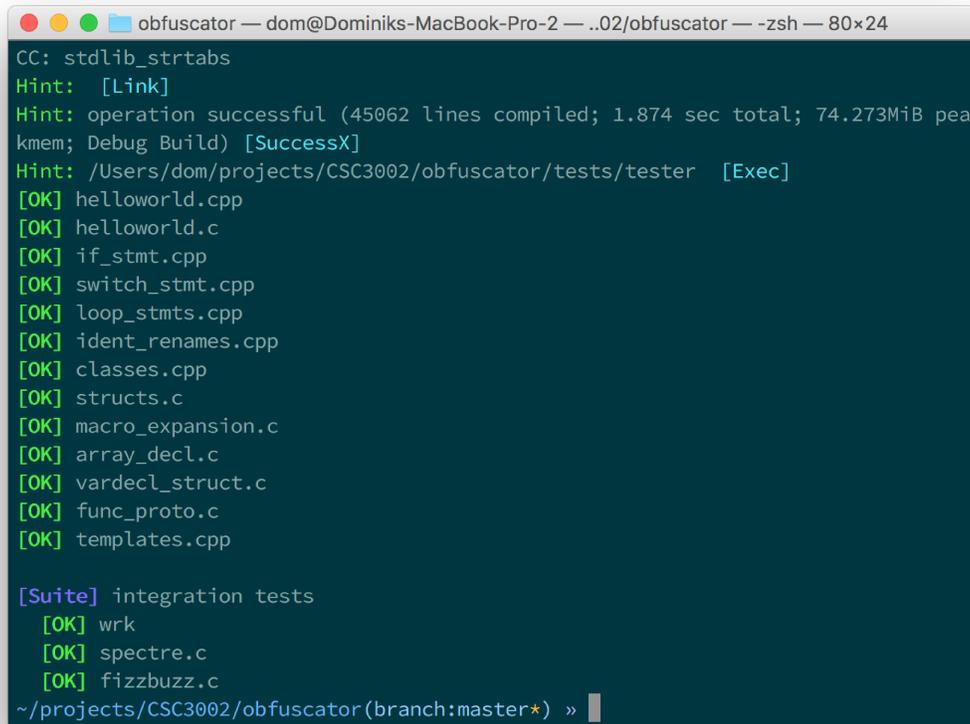

Figure 5: The test suite

For most of the unit tests, the following actions are performed:

- Source code is compiled and the resulting program executed. The output of the program is saved.
- The `obfuscator` is executed on the source code.
- Obfuscated source is compiled and the resulting program executed. The output of the program is checked against the saved output, the test passes if the outputs match.

For some unit tests additional checks are performed. For example, the `ident_renames` unit test verifies that identifiers have been renamed in the obfuscated source code.

These tests ensure that the code semantics remain the same after obfuscation and that the obfuscated code is valid C or C++ code.

Integration tests feature a similar set of actions, but for larger source code files including a large open source C project called `wrk`[10]. The `wrk` program is a high performance HTTP benchmark tool. The test suite verifies that it can be obfuscated and that the obfuscated code can be compiled.

## 6.6   Demonstration

For demonstration purposes a simple web application was put together ahead of the demo day. This web app wasn't a part of the original system requirements, it was created simply to show off the obfuscator in a more user friendly manner. Figure 6 shows what the interface of this web application looks like. It can be accessed at the following URL: https://picheta.me/obfuscator.

---

[10]https://github.com/wg/wrk

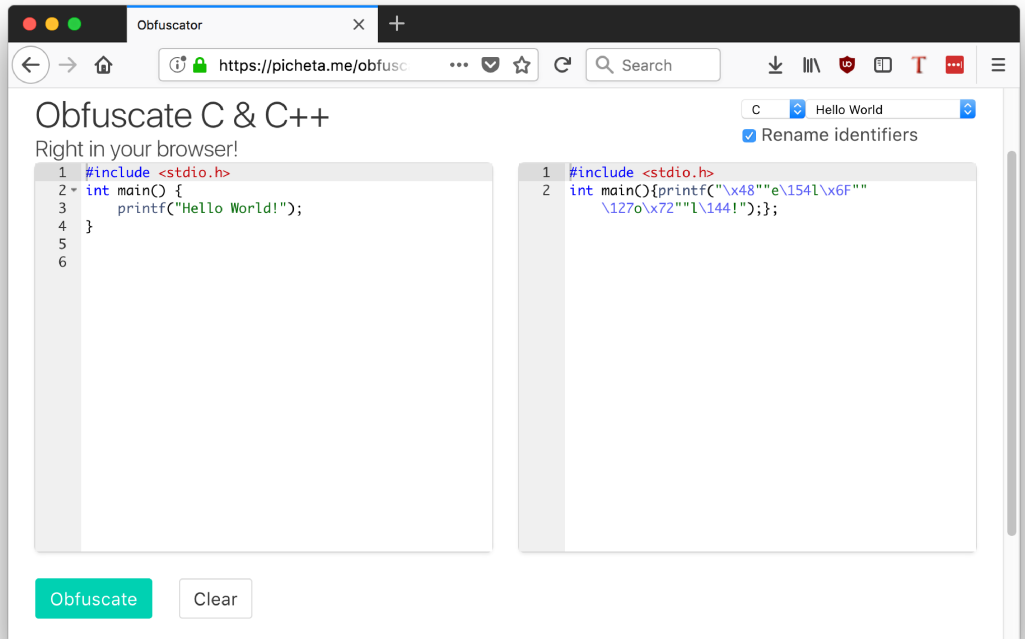

Figure 6: The Obfuscator web application

The Nim source code for the web application is available in the `web` directory of the Git repository. It's hosted on a Linux VPS and operates by executing the `obfuscator` application every time a user presses the "Obfuscate" button. It can be used as a quick and easy way to play around with the obfuscator. For more advanced usage the `obfuscator` application should be used directly.

# 7   System Evaluation and Experimental Results

In subsubsection 3.1.5, a novel approach was proposed to evaluate the `obfuscator` application as well as other similar obfuscators. The methodology is particularly well suited to evaluating the quality of a source-to-source obfuscator, because it replicates steps that a human would take to reverse engineer source code.

The evaluation will focus on 5 previously unseen C source code files. The methodology for performing this evaluation will consist of the following steps for each source code file:

- The original source code will be prettified to normalize it.
- The character count of the prettified source code will be collected.
- An obfuscator will be executed on the prettified source code.
- The character count of the obfuscated code will be collected.
- The obfuscated code will be prettified using the same settings as the original source code.
- A diff tool will be used to generate a count of added and removed characters, between the prettified original source code and the prettified obfuscated source code.

A "diff" percentage will then be calculated by comparing the number of added and removed characters in the diff, to the number of total characters in the diff.

This methodology is codified inside the `evaluate.sh` script to ensure the results can be replicated easily for each source code file.

The difference metrics should give a good indication of how resilient the obfuscator is. It will also give an indication of the obscurity. The higher the difference between the original code and the prettified code the stronger the obscurity and resilience.

## 7.1 Tools and resources

The obfuscators used in the evaluation are:

- The `obfuscator` application developed in this dissertation, with identifier renaming enabled
- Stunnix C/C++ Obfuscator (Evaluation version), with default settings
- Tigress C Obfuscator, with the `EncodeLiterals` and `EncodeArithmetic` transformations enabled on all functions

Character counts are collected using the `wc` tool. Diffing of files is performed using `git diff`.

The 5 source code files include `revcomp.gcc`, `fannkuchredux.gcc`, `regexredux.gcc-4.gcc`, `pidigits.gcc`, and `mandelbrot.gcc` from the Debian Benchmarks Game[11], these files can be downloaded from their website[12].

## 7.2 Results

Table 2: Evaluation results.

| File | Original size | Obfuscator Diff | Stunnix Diff | Tigress Diff |
|------|---------------|-----------------|--------------|--------------|
| `revcomp.gcc` | 6549 | 66.2% | 54.8% | 72.3% |
| `fannkuchredux.gcc` | 1605 | 71.5% | 57.9% | 68.8% |
| `regexredux.gcc-4.gcc` | 7106 | 58.8% | 54.9% | Error |
| `pidigits.gcc` | 1219 | 66.4% | 55.3% | 80.0% |
| `mandelbrot.gcc` | 2465 | 67.8% | 56.7% | Error |

All obfuscators perform admirably, achieving a difference metric of at least 55%.

The Tigress obfuscator was not able to obfuscate all the source code files, the failures were due to limitations in its parser which is based on CIL. Despite this it achieved some of the best metrics, reaching as high as 80% when obfuscating `pidigits`. The Tigress obfuscator is very customizable, and offers far more advanced obfuscation transformations than the ones used in this evaluation, including JIT[13] and Virtualization[14] transformations. These transformations have not been enabled as they simply wouldn't be comparable.

The results are fairly consistent for the Stunnix obfuscator, with an average of 55.9%, this is very low in comparison to the other obfuscators. It's important to note that the Stunnix obfuscator used for this evaluation was a trial version which has some limits, in particular instead of renaming identifiers completely it only prefixes them with `ReplacementFor_`. It's likely this is the explanation for its low performance.

---

[11]https://benchmarksgame-team.pages.debian.net/benchmarksgame/

[12]http://benchmarksgame.alioth.debian.org/download/benchmarksgame-sourcecode.zip

[13]http://tigress.cs.arizona.edu/transformPage/docs/jitter/index.html

[14]http://tigress.cs.arizona.edu/transformPage/docs/virtualize/index.html

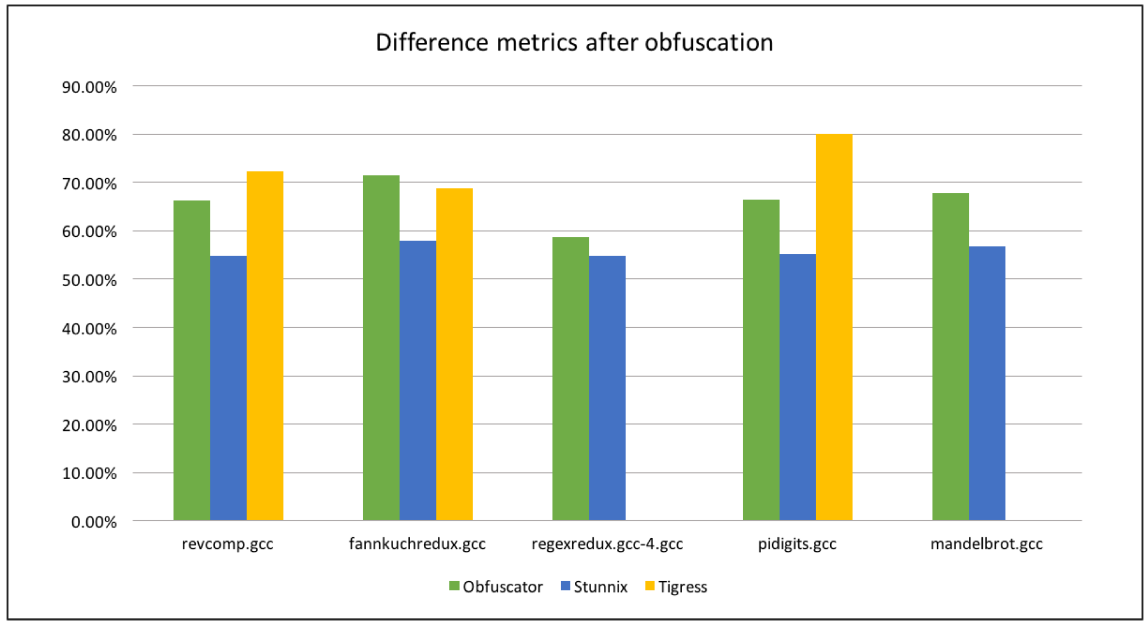

Figure 7: Chart showing the evaluation results

The results for the `obfuscator` application are far more varied, with the highest value at 71.5% and the lowest at 58.8%. It's exciting to see such great results, outcompeting a commercial obfuscator is a great feat, even if the version of Stunnix tested is limited. The `obfuscator` also outperforms Tigress for the `fannkuchredux` file, although in general Tigress performs better. But Tigress does also suffer from some parsing issues which do not affect the `obfuscator`.

In general these results show that the `obfuscator` achieves very good obscurity and resilience and that it is competitive with some of the existing C and C++ obfuscators. Additionally the `obfuscator` has been proven to be versatile, it can parse C source code that some of the leading C obfuscators can't parse. This is thanks to the clang parser it is based on.

# 8 Conclusion

As set out in the introduction, the development of a standalone cross-platform C/C++ obfuscator was pursued. The challenges primarily involved researching how to best develop the foundations for an obfuscator, including the parsing and rendering of C/C++ code. Several simple and advanced transformations have been implemented successfully as set out in the system requirements, and at least one large software project was successfully obfuscated using the developed tool. Furthermore, a novel approach to the evaluation of obfuscators was devised and used to compare the system developed to a state of the art commercial C/C++ obfuscator and to the Tigress obfuscator, yielding great results in favour of the developed system.

During development and testing several opportunities for future work have been identified. First of all, it was found that the preprocessor absorbs some vital information such as conditional compilation constructs. These are required for the appropriate obfuscation of platform-independent code. Multiple approaches to resolving this have been proposed in the relevant sections.

In addition to the above, the obfuscator currently only implements one advanced form of obfuscation. There is a lot of room for different transformations to be implemented, including ones described in detail in various research papers. Indeed, most of the time spent on this system was to research and develop the foundations for an obfuscator, the system is a good base for further obfuscation research.

Taking inspiration from a paper by Regano et al. [12], in order to properly evaluate the obfuscation quality, it would be good to create an experiment where humans attempt to reverse engineer source code under lab conditions.

Some of the design choices used in this dissertation were particularly good, including the decision to store the original code of each AST and use it to bootstrap the renderer. This allowed the obfuscator to be rapidly developed. Others were not so good, the decision to use `libclang` means that some information about the AST is not easily accessible, it provided a quick way to get started but a rewrite of this project would likely benefit from using the clang parser directly.